\documentclass[11pt]{article}

\usepackage[utf8]{inputenc}
\usepackage[T1]{fontenc}
\usepackage[english]{babel}
\usepackage{bm}

\usepackage[a4paper, margin=2.5cm]{geometry}

\usepackage{amsmath, amssymb, amsfonts}
\usepackage{graphicx}
\usepackage{hyperref}
\usepackage{authblk} 

\usepackage[numbers]{natbib}

\title{Relationship between unpredictability and intermittency in shell models of turbulence and experiments}

\author[1,2]{Ewen Frogé\thanks{Corresponding author: \texttt{ewen.froge@imt-atlantique.fr}}}
\author[1,2]{Carlos Granero-Belinchón}
\author[3]{Sébastien G. Roux}
\author[1,2]{Thierry Chonavel}
\author[3]{Nicolas B. Garnier}

\affil[1]{Department of Mathematical and Electrical Engineering, IMT Atlantique, Lab-STICC, UMR CNRS 6285,\\
655 Av. du Technopôle, Plouzané, 29280, Bretagne, France}
\affil[2]{Odyssey, Inria/IMT Atlantique, 263 Av. Général Leclerc, Rennes, 35042, Bretagne, France}
\affil[3]{ENS de Lyon, CNRS, LPENSL, UMR5672, 69342, Lyon Cedex 07, France}

\date{\today}

\begin{document}

\maketitle

\begin{abstract}
We study the predictability of turbulent velocity signals using probabilistic analog-forecasting. Here, predictability is defined by the accuracy of forecasts and the associated uncertainties. We study the Gledzer–Ohkitani–Yamada (GOY) shell model of turbulence as well as experimental measurements from a fully developed turbulent flow. In both cases, we identify the extreme values of velocity at small scales as localized unpredictable events that lead to a loss of predictability: worse mean predictions and increase of their uncertainties. The GOY model, with its explicit scale separation, allows to evaluate the prediction performance at individual scales, and so to better relate the intensity of extreme events and the loss of forecast performance. Results show that predictability decreases systematically from large to small scales. These findings establish a statistical connection between predictability loss across scales and intermittency in turbulent flows.
\end{abstract}

\section{Fully developed turbulence, intermittency and predictability}

Fully developed turbulence is a highly nonlinear and multiscale phenomenon in which energy injected at large scales is transferred through a cascade process to smaller scales, where it is dissipated by viscosity. This cascade leads to the emergence of statistical scale invariance across an intermediate range of scales $\eta<l<L$, known as the inertial range. The scales smaller than the Kolmogorov scale $\eta$ constitute the dissipative domain and the scales larger than the integral scale $L$ define the integral range. The current multifractal description of turbulence states that structure functions $S_p$ behave as:

\begin{equation}
    S_p(l) = \langle |\delta_l \mathsf{v}(x)|^p \rangle \sim l^{\zeta(p)} \, \, \, \, \text{for} \,\,\, \eta<l<L.
\end{equation}
\noindent where $\delta_l \mathsf{v}(x) = \mathsf{v}(x+l) - \mathsf{v}(x)$ is the velocity increment with $l$ denoting the scale, $\left\langle \, \right\rangle$ is the spatial average on $x$, and $\zeta(p)$ is the scaling exponent function. 

In Kolmogorov's 1941 (K41) theory, these exponents are linear: $\zeta(p) = p/3$, corresponding to a self-similar velocity field~\cite{kolmogorovLocalStructureTurbulence1941a,frischTurbulenceLegacyKolmogorov1995b}. However, empirical observations showed that $\zeta(p)$ is a concave nonlinear function of $p$~\cite{anselmetHighorderVelocityStructure1984, gagneNewUniversalScaling1990}. This is known as anomalous scaling and is associated with intermittency, \textit{i.e.} the irregular occurrence of intense, localized fluctuations of the kinetic energy dissipation~\cite{obukhovSpecificFeaturesAtmospheric1962, kolmogorovRefinementPreviousHypotheses1962}. These intense fluctuations lead to extreme events in the velocity field at small scales, both in the inertial and dissipative domains~\cite{chevillardPhenomenologicalTheoryEulerian2012} that strongly impact high-order statistics.

The multifractal formalism describes the roughness of the velocity field by considering that at each location of the flow, the velocity increment behaves as $|\delta_l \mathsf{v}(x)| \sim l^{h(x)}$, where $h(x)$ is the Hölder exponent characterizing the local regularity or roughness of the field~\cite{frischSingularityStructureFully1985,paladinAnomalousScalingLaws1987, frischTurbulenceLegacyKolmogorov1995b}. The Hölder exponent also characterizes the order of the singularity at each spatial location. Contrary to the K41 theory which assumes a self-similar turbulent velocity field with  $h=1/3$ everywhere, the multifractal models consider that the turbulent velocity field contains a variety of local singularities of different orders. The set of points with the same $h$ is fractal, and its Hausdorff dimension is given by the singularity spectrum $D(h)$\cite{frischSingularityStructureFully1985}. Having singularities of different orders is another signature of intermittency. The statistical properties of turbulence, including the structure functions, can then be derived from the distribution of local Hölder exponents\cite{frischTurbulenceLegacyKolmogorov1995b}: 

\begin{equation} 
\zeta(p) = \inf_h \left[ph + d - D(h)\right] 
\end{equation} 

\noindent where $d$ is the dimension. This formulation illustrates the relationship between anomalous scaling and the singularity distribution.

In this context, intermittency reflects the heterogeneity of the roughness of the flow, with regions of small $h$ that correspond to intense velocity gradients, and regions of mild gradients associated with large $h$. Each region contributes differently to the overall statistics. This heterogeneity explains the breakdown of self-similarity and the non-Gaussianity of velocity increments at small scales: as the scale decreases, the probability distribution of $\delta_l v$ exhibits heavy tails, reflecting the increasing contribution of rare, intense events. These features ---anomalous scaling, heavy-tailed increment distributions, and spatial heterogeneity--- are signatures of intermittency that can be described through the multifractal formalism.

Forecasting in turbulent flows, and in particular the prediction of extreme events in velocity gradients remains an active area of investigation~\cite{Ruffenach2025, bloniganAreExtremeDissipation2019, vela-martinPredictabilityIsotropicTurbulence2024}. Previous works investigated Lyapunov exponents to connect intermittency and predictability, both in dynamical shell models~\cite{crisantiIntermittencyPredictabilityTurbulence1993,aurellPredictabilitySystemsMany1996} and in direct numerical simulations of homogeneous and isotropic turbulence~\cite{boffettaChaosPredictabilityHomogeneousIsotropic2017}, highlighting how intense small-scale fluctuations can influence large-scale predictability. In a recent paper, we explored the link between unpredictability and intermittency with a data-driven approach based on analog probabilistic forecasting applied to fully developed turbulent experimental measurements from the Modane wind tunnel~\cite{frogeAnalogbasedForecastingTurbulent2025a} and showed that strong velocity fluctuations are linked to a loss of forecast accuracy.

In the present study, we extend this analysis to a simplified dynamical model of turbulence using the Gledzer–Ohkitani–Yamada (GOY) shell model.~\cite{yamadaLyapunovSpectrumChaotic1987b} Using well-separated scales by construction, this model provides a setting in which scale-specific predictability can be analyzed. This separation allows us to better understand the results observed in experimental data, particularly the relationship between intermittency and unpredictability.

We perform analog forecasting on (1) individual GOY shell variables, (2) a pseudo-velocity signal reconstructed from the sum of all GOY signals, and (3) the experimental velocity. In all cases, analog forecasting provides a probability density function (pdf) of possible future states conditioned on the past that we consider as the full probabilistic prediction. In this work, we focus on the study of its mean and variance and we illustrate their dependence on the magnitude of local fluctuations: localized high-amplitude increments are associated with worst predictions, indicating that extreme events cause larger forecasting errors in terms of both incorrect mean prediction and increase of variance. Moreover, when studying individual GOY shell variables, the estimated variance of the prediction, that we call here the uncertainty of the prediction, increases with the GOY index, \textit{i.e.} when the scale decreases. This trend highlights that small scales are also the most unpredictable, with intermittency and nonlinear interactions being more important. The statistical scarcity and intensity of extreme events dominate the prediction errors.

Very importantly, the variance of the prediction is estimated independently of the values we aim to forecast, and only depends on the pertinence of 1) the found ensemble of analogs and 2) the forecast model on this ensemble. So, localized peaks of this variance imply regions where the analogs are not behaving in a similar way. The simultaneity of the peaks of the forecast variance and the extreme velocity gradients suggests the existence of precursors of these extreme events.

The paper is organized as follows. Section II introduces the GOY shell model and the experimental turbulent velocity. Section III presents the analog forecasting method. Section IV reports results for GOY variables, pseudo-velocity, and experimental data. Section V summarizes the findings and outlines future directions.

\section{Shell models and experimental turbulent velocity}
\label{sec:Intermittency}

\subsection{The GOY shell Model}
\label{subsec:GOY}

Shell models provide a simplified representation of turbulence inspired by the Fourier representation of Navier-Stokes equations. They are dynamical models of turbulence consisting of a finite set of coupled ordinary differential equations, each equation describing the dynamics of the turbulent velocity on a shell of wavenumbers.~\cite{divletsenTurbulenceShellModels2010,biferaleSHELLMODELSENERGY2003,constantinAnalyticStudyShell2006,obukhovSpecificFeaturesAtmospheric1962} The shells, or scales, represent eddies of different sizes and are chosen to have logarithmically spaced wavenumbers. 
Coupling between neighboring shells is based on nonlinear triadic interactions. Shell models are built to reproduce important properties such as the nonlinear energy transfer between neighboring scales and intermittency, while enforcing conservation of global quantities such as energy and helicity.

\begin{figure}[t]
    \centering
    \includegraphics[width=8cm]{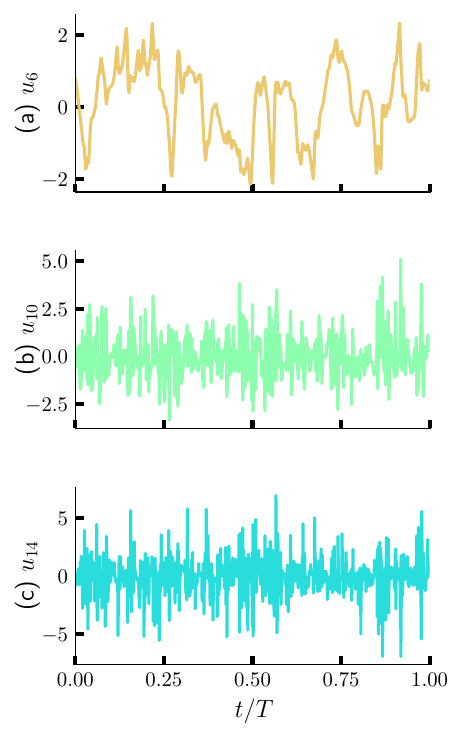}
    \caption{Normalized time series of the real part $u_i$ of the GOY complex variables a) near integral $u_6$, b) mid inertial $u_{10}$, and c) near dissipative $u_{14}$.}
    \label{fig:v_GOYmodes}
\end{figure}

The Gledzer-Ohkitani-Yamada (GOY) model~\cite{yamadaLyapunovSpectrumChaotic1987b} consists of a hierarchy of complex variables $\tilde{u}_n$, each associated with a wavenumber in the form:

\begin{equation}\label{eq:GOY}
    k_n = k_0 \lambda^n, \, \, n \in \left\lbrace 0,...,N-1 \right\rbrace \,\, n\in \mathbb{N},
\end{equation}

where $k_0$ is the smallest wavenumber, $\lambda$ sets the ratio between adjacent shells and $N$ is the total number of shells. A larger index $n$ corresponds to a larger wavenumber and hence a smaller scale. The time evolution of the system is governed by the following differential equations:

\begin{eqnarray}
    \frac{d \tilde{u}_n}{dt} =& i k_n &\left( a \tilde{u}_{n+1}^* \tilde{u}_{n+2}^* + b \tilde{u}_{n-1}^* \tilde{u}_{n+1}^* + c \tilde{u}_{n-2}^* \tilde{u}_{n-1}^* \right) - \nonumber \\
    &- \nu& k_n^2 \tilde{u}_n + f_n, 
\end{eqnarray}

where $\tilde{u}_n \in \mathbb{C}$, the asterisk indicates the complex conjugate, and $a = 1.0$, $b = -\frac{\alpha}{\lambda}$ and $c = -\frac{(1 - \alpha)}{\lambda^2}$ are constants that depend on the parameter $\alpha$ and are chosen to enforce energy and helicity conservation. These parameters together with Eq.\eqref{eq:GOY} control the nonlinear energy transfer between shells. The viscosity $\nu$ dissipates energy at small scales and $f_n$ is an external forcing term applied at large scales.

The GOY model reproduces the Kolmogorov scale invariance for shells in the inertial range. So, the structure functions follow the scaling:

\begin{equation}
    S_p(k_n) = \left\langle |\tilde{u}_n|^p\right\rangle \sim  A_p  k_n^{-\zeta(p)},
\end{equation}

\noindent where $A_p$ is a constant and $\zeta(p)\sim p/3$ up to weak non-linear corrections. Consequently, the GOY model recovers intermittency, with extreme events in both inertial and dissipative domains.

For simulations, we used the following parameters:

\begin{table}[h]
    \centering
    \begin{tabular}{ll}
    \textbf{Parameter} & \textbf{Value} \\
    \hline
    Number of shells & $N = 22$ \\
    Initial wavenumber & $k_0 = 0.125$ \\
    Scale ratio & $\lambda = 2.0$ \\
    Viscosity & $\nu = 10^{-7}$ \\
    Nonlinear parameter & $\alpha = 0.5$ \\
    Forcing strength & $f_4 = 0.005$, $f_i = 0 \quad \forall i \neq 4$ \\
    Simulation time step & $\mathrm{d}t_{sim} = 10^{-5}$ \\ 
    Sampling time step & $\mathrm{d}t = 10^{-3}$ \\
    Total simulation time & $T_{\text{max}} = 10^5$ \\
    \end{tabular}
    \caption{Simulation parameters used in the GOY model.}
    \label{tab:parameters}
    \end{table}

These parameters were chosen to ensure a developed inertial range. Forcing at $n=4$ injects energy at large scales, maintaining a statistically steady state through a cascade to smaller scales. We concentrate our study on the steady state to ensure statistical stationarity, so first tenth of the total simulation duration is discarded.

For simplicity, we restrict our analysis to the real parts of $\tilde{u}_n$. This choice is motivated by the statistical isotropy and homogeneity of the GOY model in wavenumber space, which ensure that the real and imaginary components of shell variables are statistically equivalent when averaged over long times.  In particular, their power spectra and structure functions are indistinguishable when averaged over sufficiently long trajectories. Throughout this work, $u_n=\Re(\tilde{u}_n)$ will denote the real part of the shell variable. Figure~\ref{fig:v_GOYmodes} shows the time evolution of $u_6$, $u_{10}$ and $u_{14}$ which correspond to slow, intermediate and fast dynamics, respectively. 

\subsection{Construction of a multiscale field}

The different shells of the GOY model are supposed to represent disjoint ensembles of wavenumbers in the Fourier space. Thanks to this disjointness and the linearity of the Fourier transform, we can define a multiscale field by summing the $u_n$:

\begin{equation}\label{eq:pseudov}
    \mathsf{v}_{\text{GOY}}(t) = \sum_{n=0}^{N} u_n(t).
\end{equation}

\begin{figure}[t]
    \centering
    \includegraphics[width=8cm]{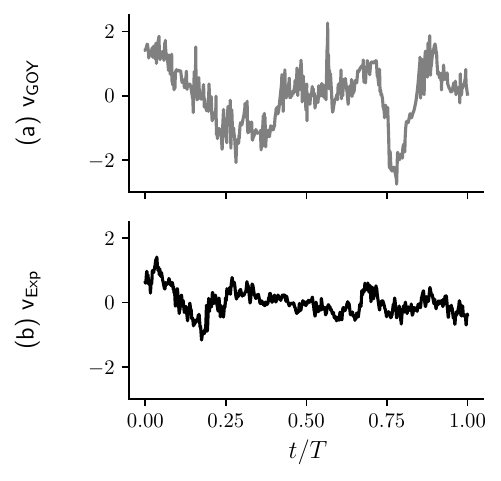}
    \caption{Normalized time series of a) the GOY pseudo-velocity $\mathsf{v}_{\text{GOY}}(t)$, b) the experimental velocity $\mathsf{v}_{\text{Exp}}(t)$ from Modane wind tunnel.}
    \label{fig:v_GOYsum_Modane}
\end{figure}

\noindent This definition reconstructs a multiscale signal that integrates contributions from all scales, see Appendix~\ref{sec:appendix}. We call this process GOY pseudo-velocity. Figure~\ref{fig:v_GOYsum_Modane} a) illustrates the time evolution of this GOY pseudo-velocity which presents multiscale dynamics.

\begin{figure}[ht]
    \centering
    \includegraphics[width=8cm]{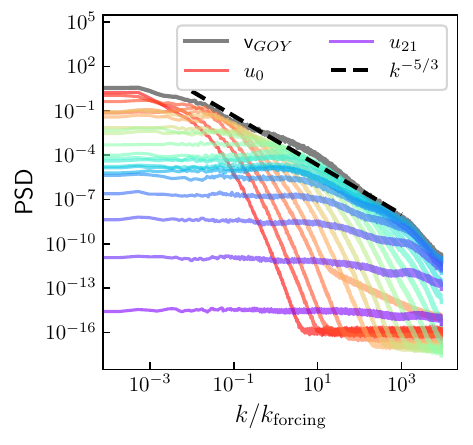}
    \caption{Power spectral density (PSD) of individual $u_n$ (colored lines) and of the pseudo-velocity (gray line). The PSD is plotted against a surrogate wavenumber axis $k$, normalized by the forcing scale $k_\text{forcing} $. The theoretical $ k^{-5/3} $ inertial range scaling is shown for reference (dashed black line).}
    \label{fig:spectrum_goy}
\end{figure}

Figure~\ref{fig:spectrum_goy} presents the power spectral density (PSD) of $u_n$ (colored lines) and the summed pseudo-velocity (gray line). Using the eddy turnover time $\tau(k) \sim (k \mathsf{v}_k)^{-1}$ and the characteristic velocity at scale $k$, $\mathsf{v}_k \sim k^{-1/3}$, temporal frequencies $\omega$ are mapped to an effective wavenumber via $ k\sim \omega^{3/2}$, see~\cite{divletsenTurbulenceShellModels2010}. This allows the power spectral density to be plotted on a wavenumber-like axis and compared to spatial spectra from experimental turbulence. The resulting spectrum of the pseudo-velocity constructed from all shells exhibits a scaling close to $k^{-5/3}$ over a finite range of scales, consistent with Kolmogorov's inertial range scaling. This illustrates the multiscale nature of the process defined in Eq.(\ref{eq:pseudov}).

\subsection{Experimental turbulent velocity}

Turbulent velocity was measured via hot-wire anemometry in a turbulent flow in the Modane wind tunnel~\cite{kahalerrasIntermittencyReynoldsNumber1998a}. With a Taylor-scale Reynolds number $R_\lambda \approx 2500$, the flow exhibits fully developed turbulence. Using Taylor's frozen field hypothesis~\cite{taylorSpectrumTurbulence1938}, temporal variations can be interpreted as spatial variations advected by the mean flow. This gives access to the spatial fluctuations of turbulent velocity from single-point Eulerian velocity measurements. The sampling frequency of the experiment is $f_s = \frac{1}{dt} = 25$ kHz and allows to probe scales smaller than the Kolmogorov scale that is of the order of $5dt$.  Figure~\ref{fig:v_GOYsum_Modane} b) presents the time evolution of the experimental turbulent velocity  $\mathsf{v}_{\rm exp}$ over one integral scale $T=2350dt$.

\section{Forecast \& Analogs}
\label{sec:method}

\subsection{Probabilistic forecast and Innovation}

For a stochastic process $X=\{x_t\}$ uniformly sampled at $dt$, future values $x_t$ of $X$ can be derived statistically from previous samples $x_{-\infty:t-dt}$. A probability density function of possible future states conditioned on the past, $q(x_t | x_{-\infty:t-dt})$, can be estimated. In this framework, the mean prediction and its uncertainty are given by the expected value of $x_t$ conditioned on its complete past $\hat{x}_t=\mathbb{E}[x_t | x_{-\infty:t-dt}]$ and the variance of $x_t$ conditioned on its complete past $\sigma_t^2=\mathbb{V}[x_t | x_{-\infty:t-dt}]$. Consequently, the probabilistic prediction is the full pdf $q(x_t | x_{-\infty:t-dt})$. In this work, we will study it through its mean $\hat{x}_t$ and variance $\sigma_t^2$ and for simplicity of language we call $\hat{x}_t$ the prediction and $\sigma_t^2$ its uncertainty.

The innovation $\epsilon_t$ quantifies the unpredictable component of $x_t$. It is defined as:

\begin{equation}
    \epsilon_t = x_t - \mathbb{E}[x_t | x_{-\infty:t-dt}]
\end{equation}

In practice, conditioning is restricted to a finite past window of length $p$, leading to the approximations:

\begin{eqnarray}
    \mathbb{E}[x_t | x_{t-pdt:t-dt}] &=& \int x_t q(x_t | x_{t-pdt:t-dt}) dx_t \\
    \mathbb{V}[x_t | x_{t-pdt:t-dt}] &=& \int x_t^2 q(x_t | x_{t-pdt:t-dt}) dx_t \nonumber \\
    & & - \left( \int x_t q(x_t | x_{t-pdt:t-dt}) dx_t \right)^2 \\
    \epsilon_t^{(p)} &=& x_t - \mathbb{E}[x_t | x_{t-pdt:t-dt}]
\end{eqnarray}

\subsection{Analog-Based Estimation}

For a given time $t$, we consider the sequence of size $p$ preceding $x_t$:  

\begin{equation}
    \vec{x}_{t}^{(p)} =
    \begin{pmatrix}
        x_{t-dt} \\
        \vdots \\
        x_{t-p dt}
    \end{pmatrix}
\end{equation}

We say that $x_t$ is the successor of $\vec{x}_{t}^{(p)}$. Two sequences $\vec{x}_{t}^{(p)}$ and $\vec{x}_{t'}^{(p)}$ are analogs if they are close in terms of a given norm~\cite{lorenzAtmosphericPredictabilityRevealed1969a}. Based on Poincar\'e’s recurrence theorem, we assume that in a sufficiently long time-series, close analogs have similar successors\cite{poincareProblemeTroisCorps1890a}. Consequently, for a given sequence $\vec{x}_{t}^{(p)}$, it is possible to perform a statistical prediction of $x_t$ based on regressions between its past analogs and their successors. This approach yields an estimate of the conditional probability density function $q^{(p)}(x_t | x_{t-pdt:t-dt})$ and in particular its mean and variance.

Given $\vec{x}_{t}^{(p)}$, the $k$ closest historical sequences $\{\vec{x}_{t_i}^{(p)} \}_{1\le i\le k}$ with $t_i<t$ serve as analogs~\cite{platzerUsingLocalDynamics2021c,lguensatAnalogDataAssimilation2017a}. Following~\cite{frogeAnalogbasedForecastingTurbulent2025a}, in order to estimate $\mathbb{E}[x_t | x_{t-pdt:t-dt}]$ and $\mathbb{V}[x_t | x_{t-pdt:t-dt}]$, we use a linear regression approach, where each analog sequence is used to predict its successor. We denote by $\hat{x}_t^{(p)}$ the linear regression estimator of $x_t$ and $\hat{\sigma}_t^{(p)}$ its standard deviation. Given $k$ analogs, the regression minimizes $||\vec{y} -  \bm{X}_p \beta_t||^2_{\mathbf{W}}$ over $\beta_t$, where:

\begin{itemize}
    \item $\vec{y} = (x_{t_1'}, \dots, x_{t_k'})^T$ is the vector of analog successors,
    \item $\bm{X}_p$ is the $k \times (p+1)$ matrix of past sequences augmented with a column of ones to include a bias term,
    \item $\bm{W}$ is a diagonal weight matrix.
\end{itemize}

The least-squares estimator is:  
\begin{equation}
    \hat{\beta}_t = (\bm{X}_p^T \bm{W} \bm{X}_p)^{-1} \bm{X}_p^T \bm{W} \vec{y},
\end{equation}
leading to:
\begin{equation} \label{eq:meanvariance}
    \hat{x}_t^{(p)} = \hat{\beta}_t^T \vec{x}_{t}^{(p)}, \quad \hat{\sigma}_t^{(p)} = \sqrt{\sum_{i=1}^{k} w_{ii} (x_{t'_i} - \hat{x}_{t'_i})^2}
\end{equation}
The weights $ w_{ii} $ are defined as
\begin{equation}
    w_{ii} = \frac{\exp\left(- d_i / m \right)}{\sum_{j=1}^{k} \exp\left(- d_j / m \right)},
\end{equation}
where $ d_i=||\vec{x}_{t}^{(p)}-\vec{x}_{t'}^{(p)}||$  is the distance to the $ i $-th analog and $ m $ is the median of all $ d_i $, giving more weight to closer analogs.

The innovation estimate is given by
\begin{equation}
    \hat{\epsilon}_t^{(p)} = x_t - \hat{x}_t^{(p)}
\end{equation}

Very importantly, while the innovation depends on the specific value $x_t$ we aim to predict, this is not the case for $\hat{\sigma}_t^{(p)}$ that is estimated as the difference between the successors and the predictions made on the analogs. Consequently, $\hat{\sigma}_t^{(p)}$ yields information of the relevance of 1) the found analogs and 2) the fitted model on this ensemble of analogs.

To focus on the relative structure of fluctuations, each sequence of size $p$ is centered prior to analog search. 
This preprocessing step ensures that similarity between sequences is evaluated based on their variation rather than their mean value~\cite{frogeAnalogbasedForecastingTurbulent2025a}.

The choice of the forecasting parameters ---the analog size $p=3$, the number of analogs $k=70$, and the size $N=2^{22}$ of the reference analog dictionary--- follows the settings established in our previous study\cite{frogeAnalogbasedForecastingTurbulent2025a}. These values were selected based on empirical performance and robustness criteria evaluated on experimental turbulence data. For GOY time series, the sampling time of the GOY model $dt=10^{-3}$ simulation units, is used as prediction time step in the analog forecasting method. For the experimental turbulent velocity, the sampling time $dt=0.4$ ms is used. In both cases, these times correspond to time scales in the dissipative domain.

\section{Results}

The GOY dataset consists of $22$ shell variables $u_n \in \mathbb{R}$. For each one, the time-series is partitioned into twelve non-overlapping subsections with $N = 2^{22}$ samples in each one. The first subsection is used as a reference database for analog searching, and the probabilistic forecast is performed on the last ten. To ensure the independence between samples used in analog search and samples used in forecasting, the second partition is not used. The same procedure is applied to the pseudo-velocity field derived constructed following Eq.\eqref{eq:pseudov}, with segments of $2^{22}$ samples, and to the experimental turbulent velocity with segments of $2^{21}$ samples. The statistics presented in the following are computed over the last ten partitions, while time series plots show only one integral scale.

\subsection{Prediction and uncertainty in GOY}

\begin{figure}[ht]
    \centering
    \includegraphics[width=8cm]{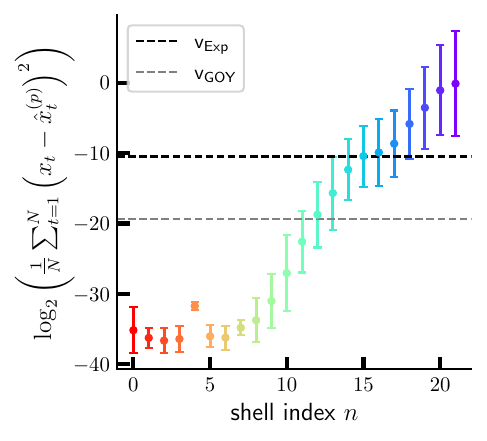}
    \caption{Logarithm of the MSE of the forecast $ \frac{1}{N} \sum_{t=1}^{N} \left( x_t - \hat{x}_t^{(p)} \right)^2$ as a function of GOY shell index. Variables are normalized prior to any computation as \( x_t \mapsto \frac{x_t - \mu}{\sigma} \), where \( \mu \) and \( \sigma \) are the mean and standard deviation, respectively.
    The color scale encodes the index for consistency with other figures.}
    \label{fig:innovation_modes}
\end{figure}

Figure~\ref{fig:innovation_modes} presents the empirical Mean Squared Error (MSE) $\frac{1}{N} \sum_{t=1}^{N} \left( \hat{\epsilon}_t^{(p)} \right)^2$ of the forecast across all $u_n$. It corresponds to the time-averaged variance of the innovation signal and characterizes the overall predictability of the shell component.

For shells in the inertial and dissipative domains, the innovation variance clearly increases with the index, \textit{i.e.} large scales are more predictable than small scales. The systematic reduction in predictability as one moves toward smaller scales is consistent with the picture of turbulence as an energy cascade where predictability is progressively lost due to the increasing role of intermittency.

\begin{figure*}[t]
    \centering
    \includegraphics[width=18cm]{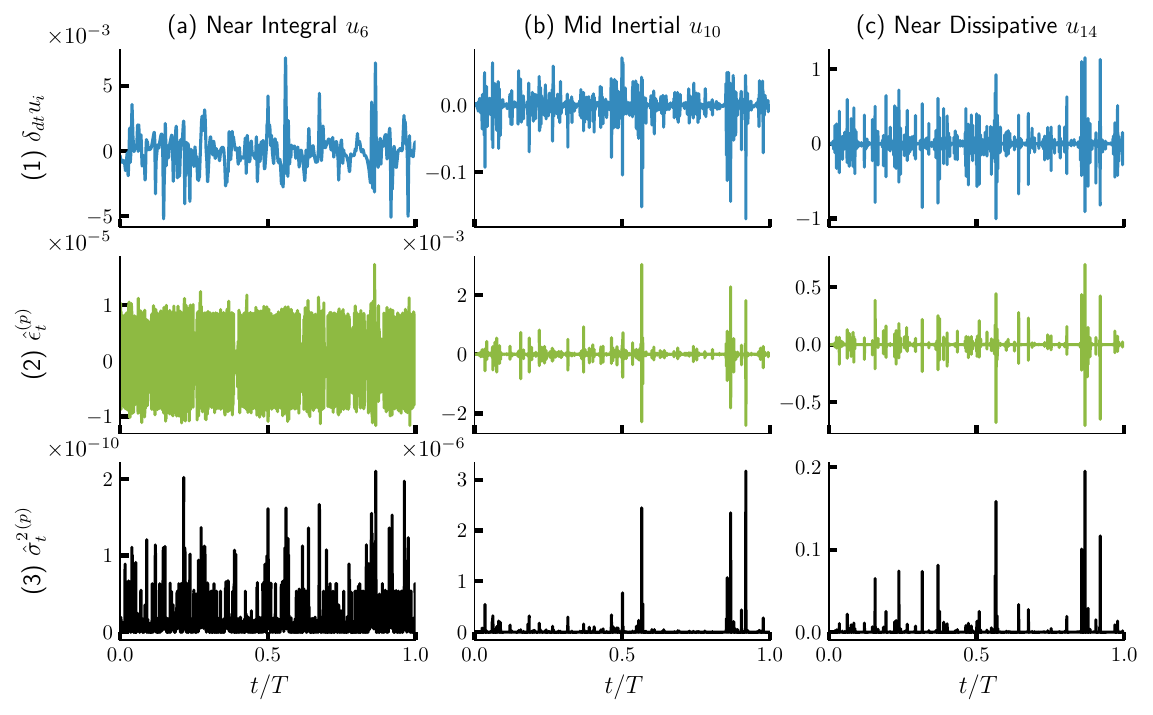}
    \caption{Time series of the increments $\delta_{dt} u_i = u_i(t+dt) - u_i(t)$ (blue), innovation $\hat{\epsilon}_t^{(p)}$ (green), and prediction variance $\hat{\sigma}_t^{2\,(p)}$ (black) for the real parts of GOY shells $6$, $10$, and $14$.}
    \label{fig:signals_goy_modes}
\end{figure*}

To better understand how predictability evolves across scales, we analyze three representative shells of the GOY model: one near the integral scale ($u_6$), one in the mid-inertial range ($u_{10}$), and one in the inertial domain close to the dissipative range ($u_{14}$). Figure~\ref{fig:signals_goy_modes} presents for the three chosen shell variables: 1) the time serie $u_i$, 2) the smallest increments $\delta_{dt} u_i$ which quantify the intensity of fluctuations over a time step $dt$, 3) the innovation $\hat{\epsilon}_t^{(p)}$ that measures the difference between the actual value and the prediction based on analog selection and 4) the prediction variance $\hat{\sigma}_t^{(p)} $ that estimates the uncertainty of the prediction. 

Shell variable $u_6$ depicts a large scale and so varies smoothly over time. The increments $\delta_{dt} u_6$ remain small and evolve slowly, reflecting the variable smoothness. Its innovation $\hat{\epsilon}_t^{(p)}$ is several orders of magnitude smaller than the increments, indicating high predictability. The prediction variance $\hat{\sigma}_t^{2\,(p)}$ is of the order of $10^{-8}$ and is therefore negligible when compared to the variance of the increment. This indicates stable predictions and confirms the high predictability at this scale. The dynamics at this scale are largely dominated by the external large-scale forcing that is smooth, resulting in structures that evolve smoothly.

The shell $10$ corresponds to an intermediate scale in the inertial domain, between the smooth evolution at larger scales and the strongly intermittent fluctuations at smaller scales in the inertial and dissipative ranges. $u_{10}$ is rougher than $u_6$, and its increment $\delta_{dt} u_{10}$ fluctuates more, as expected for the small scales of turbulence. The innovation is larger than in shell 6 but two orders of magnitude smaller than the amplitude of the increment. The prediction variance $\hat{\sigma}_t^{2\,(p)}$ is of the order of $10^{-7}$, increasing slightly compared to shell 6, but still negligible compared to the amplitude of the increment. These results suggest that while prediction is less effective than at larger scales, it still approximates future values well.

Shell variable $u_{14}$, which characterizes the smallest scales of the inertial domain, shows sharper and stronger fluctuations compared to $u_6$ and $u_{10}$, and the increment $\delta_{dt} u_{14}$ is visually more intermittent. The innovation is now of the order of the amplitude of the increment and sometime even larger. This indicates that predicting the dynamics of $u_{14}$ is complicated. The prediction variance $\hat{\sigma}^{2\,(p)}_t$ also approaches values close to the variance of the increment, illustrating the increased uncertainty in the predictions. This loss of predictability observed for $u_{14}$ is consistent with the increasing significance of extreme events at small scales, a characteristic feature of intermittency in turbulence. 

To better understand the loss of predictability, we analyze the variance of the innovation conditioned on the square of the increment:

\begin{equation}
\langle \hat{\sigma}_t^{2\,(p)} \mid (\delta_{dt} u_i)^2 \rangle = \int P(\hat{\sigma}_t^{2\,(p)} \mid (\delta_{dt} u_i)^2) \hat{\sigma}_t^{2\,(p)} \, d\hat{\sigma}_t^{2\,(p)}.
\end{equation}

\noindent where $\hat{\sigma}^{2\,(p)}_t$ is the variance of the prediction and $P(\hat{\sigma}^{2\,(p)}_t \mid (\delta_{dt} u_i)^2)$ is the distribution of variances conditioned on the values of the square of the increments. This is done via a joint histogram of \( \log[(\delta_{dt} u_i)^2] \) and \( \log[\hat{\sigma}_t^{2\,(p)}] \), from which the conditional mean is computed for each increment bin. This quantity highlights whether forecasting errors are uniformly distributed across different fluctuation magnitudes or whether extreme events contribute disproportionately to the loss of predictability.

\begin{figure}[ht] 
    \centering 
    \includegraphics[width=8cm]{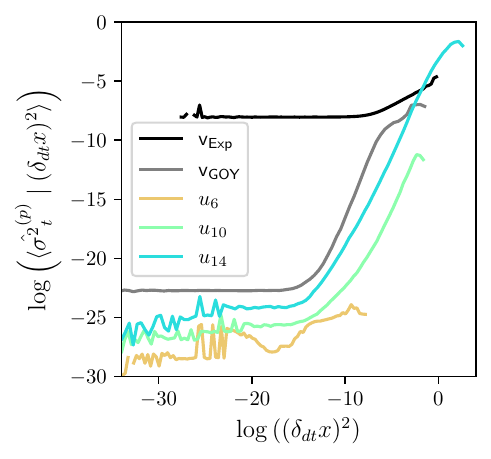} 
    \caption{Logarithm of the variance of innovation conditioned on the squared increment magnitude, $ \langle \hat{\sigma}_t^{2\,(p)} \mid (\delta_{dt} x)^2 \rangle $, for GOY $u_6$ (orange), $u_{10}$ (green), and $u_{14}$ (cyan), as well as for the pseudo velocity $\mathsf{v}_{\rm GOY}$ (grey) and the experimental turbulent velocity  $\mathsf{v}_{\rm exp}$ (black). } 
    \label{fig:varconditioned_GOY} 
\end{figure}

Figure~\ref{fig:varconditioned_GOY} presents the conditioned variance of innovation for shells $6$, $10$, and $14$. For $u_6$, the conditioned variance remains relatively stable across all increment magnitudes. Also, increments of largest amplitudes are of the order of $\log((\delta_{dt}u_i)^2)\sim -10$. This suggests that there are no extreme events in this shell variable impacting the prediction performance. Thus, large-scale dynamics are primarily driven by external forcing, leading to relatively predictable evolution. For $u_{10}$, increments with the largest amplitudes are of the order of $\log((\delta_{dt}u_i)^2)\sim -5$ and a slight increase in conditioned variance is observed for squared increment magnitude between $\log((\delta_{dt}u_i)^2)\sim -15$ and $\log((\delta_{dt}u_i)^2)\sim -5$. This suggests that while most prediction errors remain moderate, large increments lead to slightly larger forecasting errors, indicating the growing role of intermittency in shaping predictability. For $u_{14}$, the conditioned variance exhibits a steep increase with the squared increment amplitude, confirming that extreme events dominate the loss of predictability at small scales. The sharp rise in conditioned variance highlights how intermittency-driven extreme events cause unpredictability at small scales. This aligns with results of Figure~\ref{fig:innovation_modes}: at small scales extreme events are becoming increasingly frequent and intense leading to a strong loss in predictability.

\subsection{Statistical characterization of increments and innovations in GOY}

To analyze the performance of the analog-based forecasting, we estimate the autocorrelation function $R(\tau)$ of the innovation $\hat{\epsilon}_t^{(p)}$ , defined for a generic centered process $X(t)$ as $R(\tau)=\frac{\left\langle X(t)X(t+\tau) \right\rangle}{\left\langle X^2(t) \right\rangle}$, and its kurtosis $ \frac{\left\langle X^4(t) \right\rangle}{\left\langle X^2(t) \right\rangle^2}$. The autocorrelation of the innovation quantifies the remaining linear dependencies while its kurtosis characterizes the existence of rare, large magnitude prediction errors. In theory, the innovation would be a white noise, but in practice, using a finite past of length \( p \), some higher-order dependencies may remain. We compare the results to the autocorrelation function and the kurtosis of the smallest-scale increment $\delta_{dt} u_i$.

\begin{figure}[t]
    \centering
    \includegraphics[width=8cm]{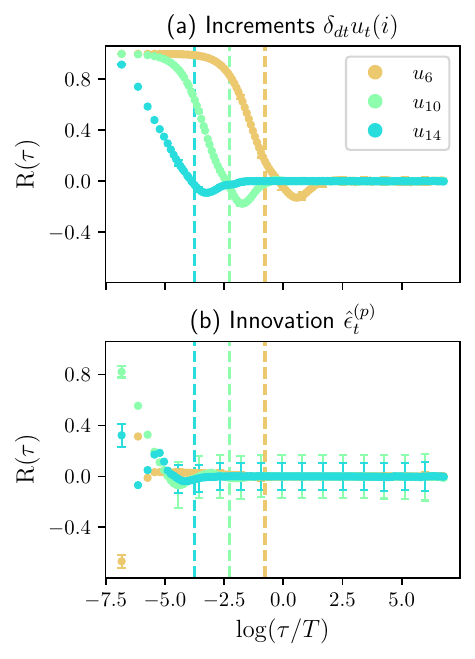}
    \caption{Normalized autocorrelation $R(\tau)$ of the increment (a) and the innovation (b) for $u_6$, $u_{10}$, and $u_{14}$ in function of $\log(\tau/T)$. Vertical dashed lines indicate the eddy turnover time of each shell variable.}
    \label{fig:R_innovation}
\end{figure}

Figure~\ref{fig:R_innovation} displays the autocorrelation of the increment (a) and the innovation (b) for $u_6$, $u_{10}$ and $u_{14}$. All shell variables present an autocorrelation of the increment that decreases and reaches zero at a time scale consistent with the eddy turnover time of the corresponding shell variable. 
Lower-index shell variables, such as $u_6$, exhibit broader ranges of positive autocorrelation, while higher indices, like $u_{14}$, lose correlation more quickly. This is consistent with faster dynamics at smaller scales. In contrast, the innovations of the shell variables show autocorrelation curves that are almost zero when we move away from the origin. This absence of correlation structure indicates that the innovation is close to a white noise, even when the increment exhibits temporal coherence. The analog forecasting procedure thus removes the linearly correlated part of the signal. 

Table~\ref{tab:kurtosis} reports the kurtosis of the innovations and the increments of $u_6$, $u_{10}$, and $u_{14}$. While the increments already display high kurtosis, increasing with GOY index and indicating non-Gaussianity and growing intermittency, the kurtosis of the innovations is substantially larger. The difference between the kurtosis of increments and innovations is stronger at small scales: for $u_{14}$, the innovation kurtosis exceeds that of increments by two orders of magnitude. These values confirm that although temporal correlations are removed, the innovation retains the non-Gaussian features of turbulence — specifically, the intermittent, burst-like events that remain beyond the reach of prediction. The innovation thus concentrates the statistically extreme components of the signal that the analog method fails to forecast.

\begin{table}[ht]
    \centering
    \caption{Logarithm of the kurtosis of increments and innovations for GOY $u_6$, $u_{10}$, and $u_{14}$, as well as for the GOY pseudo-velocity and the experimental velocity.}
    \label{tab:kurtosis}
    \begin{tabular}{c|c|c}
    \hline
    \textbf{Signal} & $ \log \mathrm{Kurt}[\delta_{dt} u_i] $ & $ \log \mathrm{Kurt}[\hat{\epsilon}_t^{(p)}] $ \\
    \hline
    $u_6$            & $2.38 \pm 0.11$ & $3.92 \pm 0.10$ \\
    $u_{10}$         & $3.46 \pm 0.07$ & $8.37 \pm 1.05$ \\
    $u_{14}$         & $4.75 \pm 0.09$ & $8.21 \pm 0.20$ \\
    $\mathsf{v}_{\text{GOY}}$ & $5.50 \pm 0.25$ & $9.23 \pm 0.38$ \\
    $\mathsf{v}_{\text{Exp}}$ & $2.24 \pm 0.03$ & $2.79 \pm 0.08$ \\
    \hline
    \end{tabular}
    \end{table}

\subsection{Prediction and uncertainty in GOY pseudo-velocity field and experimental velocity}

In this section, we evaluate the performance of analogs forecast in more complex processes: the GOY pseudo-velocity field generated from Eq.\eqref{eq:pseudov} and an experimental turbulent velocity signal. Both processes exhibit multiscale dynamics, an energy cascade, and intermittency. In both cases, we are able to link extreme events in velocity increments to unpredictable events.

\begin{figure*}[t]
    \centering
    \includegraphics[width=18cm]{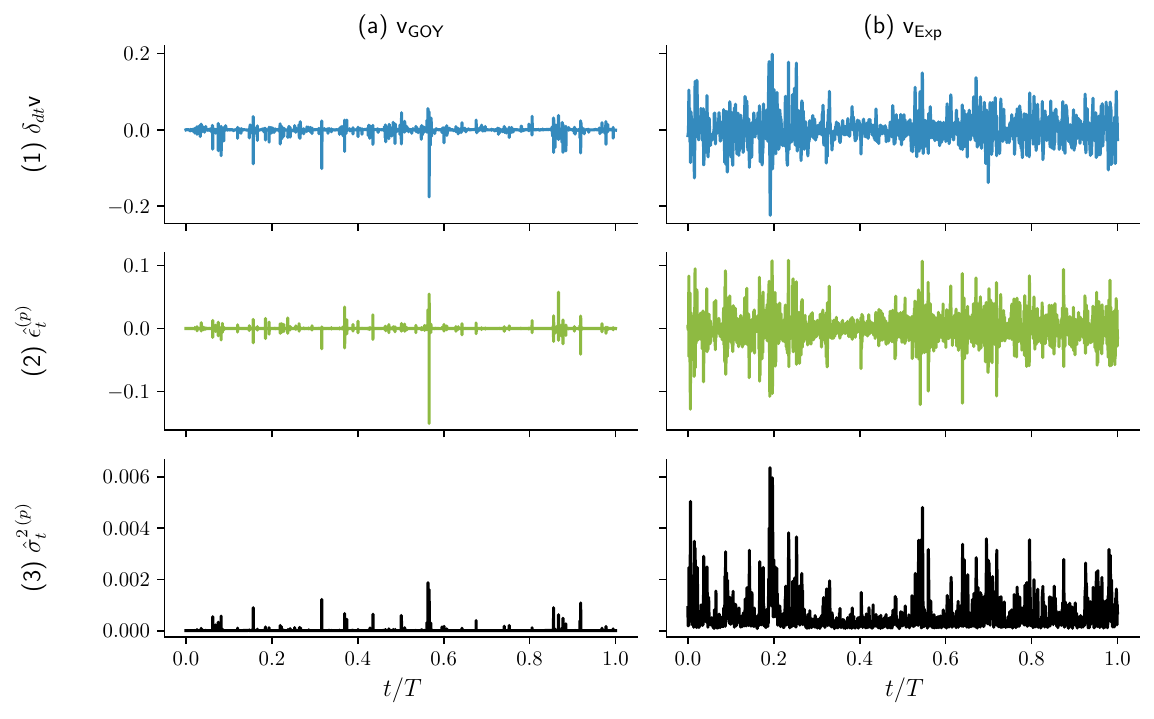}
    \caption{Time series of (a) the GOY pseudo-velocity $\mathsf{v}_{\text{GOY}}(t)$ and (b) the experimental velocity $\mathsf{v}_{\text{Exp}}(t)$ , along with their respective increment (1), innovation (2) and variance of the prediction (3). }
    \label{fig:signals_GOYsum_Modane}
\end{figure*}

In Figure~\ref{fig:signals_GOYsum_Modane}, for the GOY pseudo-velocity $\mathsf{v}_{\text{GOY}}(t)$ and for the experimental velocity $\mathsf{v}_{\text{Exp}}(t)$: , we present the smallest increment of the velocity $\delta_{dt}\mathsf{v}(t)$, the innovation $\hat{\epsilon}_t^{(p)}$ and the uncertainty of the prediction $\hat{\sigma}_t^{2\, (p)}$, all plotted over time.

The increment of the GOY pseudo-velocity exhibits lower variance but higher intermittency with very strong peaks appearing at isolated times. The increment of $\mathsf{v}_{\text{Exp}}$ is also intermittent but the ratio of the amplitudes of extreme and usual events is smaller. The same behavior is observed for the innovation. In the case of the GOY pseudo-velocity, most of the velocity values are well predicted. Most bursts in the innovation coincide with extreme events of the increment. In the case of $\mathsf{v}_{\text{Exp}}$, average prediction is less accurate and  isolated innovation peaks also correspond to unpredictable events linked to extreme events of the increment. Finally, the prediction uncertainty is also intermittent for both velocities, especially for the GOY pseudo-velocity, with strong peaks aligned with those in innovation, \textit{i.e.} extreme events of the increment lead to particularly unpredictable events with a high uncertainty in the prediction.

Figure~\ref{fig:varconditioned_GOY} shows the  prediction variance  conditioned on the squared velocity increment, for both GOY pseudo-velocity and experimental velocity. In both cases, the conditioned variance remains relatively stable for small-amplitude increments, but increases significantly for larger increments. This confirms that extreme events are the dominant contributors to predictability loss, as forecasting errors grow with the increment magnitude. Despite quantitative differences, we see that for both studied processes the predictability loss is driven mainly by rare, large-amplitude events of the velocity increment.

\begin{figure}[t]
    \centering
    \includegraphics[width=8cm]{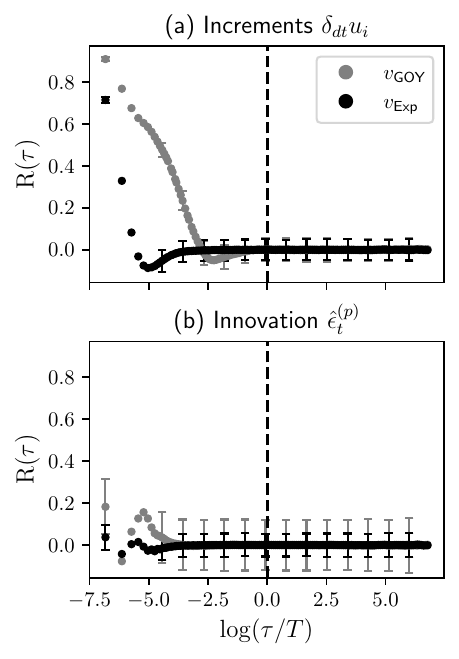}
    \caption{Normalized autocorrelation function $R(\tau)$ of the smallest-scale increment (a) and the innovation (b), for the GOY pseudo-velocity $\mathsf{v}_{\text{GOY}}$ and the experimental velocity $ \mathsf{v}_{\text{Exp}} $ . Vertical dashed lines indicate the eddy turnover time of forced shell variable $u_4$ for GOY and integral scale for experimental velocitiy.}
    \label{fig:R_innovation_sum}
\end{figure}

Figure~\ref{fig:R_innovation_sum} shows the autocorrelation function $R(\tau)$ of the increment and of the innovation for $\mathsf{v}_{\text{GOY}}$ and $\mathsf{v}_{\text{Exp}}$. In both cases, the autocorrelation of the increment decays with increasing $\tau$. For $\delta_{dt}\mathsf{v}_{\text{GOY}}$, the decay is slower and correlation persists over a wider range of time scales. The autocorrelation of $\delta_{dt}\mathsf{v}_{\text{Exp}}$ decreases more rapidly. In contrast, the autocorrelations of the innovation of both velocities remain almost flat and close to zero at all $\tau$, indicating that innovations are decorrelated. The absence of autocorrelation characterizes the innovation as a residual that does not retain linear temporal structure.

\begin{figure}[t]
    \centering
    \includegraphics[width=8cm]{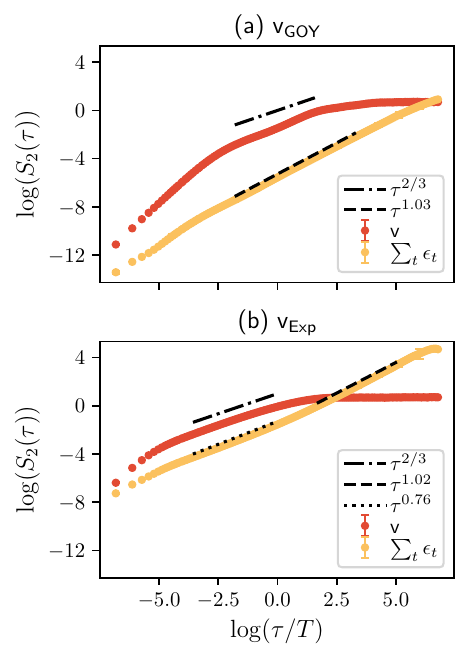}
    \caption{Second-order structure function $\log(S_2(\tau))$ of the velocity and the cumulative sum of the innovation (orange) for the GOY pseudo-velocity $\mathsf{v}_{\text{GOY}}$ (a) and experimental velocity  $\mathsf{v}_{\text{Exp}}$ (b) in function of the scale of analysis $\log(\tau/T)$. For the GOY pseudo-velocity $T$ is the eddy turnover time of forced shell variable $u_4$, and for the experimental velocity $T$ is the integral scale of the flow.}
    \label{fig:s2_f_innovation_sum}
\end{figure}

Figure~\ref{fig:s2_f_innovation_sum} shows the second-order structure function $\log(S_2(\tau))$ of the velocity (red) and the cumulative sum of the innovation (orange), for both the GOY pseudo-velocity (a) and the experimental velocity (b), as a function of the scale of analysis $\log(\tau/T)$. In both cases, the structure function of the velocity follows a scaling close to $S_2(\tau) \sim \tau^{2/3}$ in a full range of scales, consistent with Kolmogorov’s theory for fully developed turbulence. The second order structure function of the cumulative sum of the innovation shows power-law behaviors with different exponents. In the GOY case, the slope is close to 1 across all scales, which suggests that the innovation is decorrelated: the cumulative sum follows a random walk, as expected for a white-noise residual. In contrast, the cumulative sum of the innovation in the experimental velocity exhibits a two-regime behavior: at small scales, the slope is approximately 0.76, while at larger scales it transitions to a slope close to 1. This deviation from ideal white-noise behavior at small scales indicates that the innovation of experimental turbulence may retain weak correlations that are not removed by the analog forecasting method.

Table~\ref{tab:kurtosis} reports the logarithm of the kurtosis of the smallest-scale increment and of the innovation for both the GOY pseudo-velocity and the experimental velocity. The increments of the GOY signal display a high logarithm of the kurtosis of $5.50$, while the increments of $\mathsf{v}_{\text{Exp}}$ have a significantly lower value around $2.24$. This difference indicates that the GOY increments are more intermittent, with more pronounced extreme deviations from the mean. For both signals, the kurtosis of the innovation is higher than that of the increment. In the GOY pseudo-velocity, this effect is particularly strong. In the experimental signal, the increase is smaller, but still present. This suggests that in both cases, the unpredictable residual is characterized by heavy-tailed statistics and is dominated by rare, large-amplitude deviations.

\section{Conclusion}

In this study, we analyzed the scale-dependent structure of predictability in turbulent signals using analog-based probabilistic forecasting. We studied three different systems: individual shells of the Gledzer–Ohkitani–Yamada (GOY) model, a GOY pseudo-velocity defined as the sum of all the GOY shell variables, and a high-Reynolds number experimental velocity signal from the Modane wind tunnel.

The results from the individual GOY shell variables reveal a clear increase in the mean prediction error as the scale decreases. In addition, localized extreme events in the increment of the shell variables correspond to high values of the innovation and the prediction uncertainty. This trend is also observed in the GOY pseudo-velocity and in the experimental velocity. This behavior shows that strong local fluctuations linked to intermittent events are harder to predict, and contribute significantly to the overall forecasting error.

To further characterize the statistical structure of the innovation, we analyzed its autocorrelation and kurtosis. For $u_6$, $u_{10}$, and $u_{14}$ as well as for the GOY pseudo-velocity and the experimental velocity, the autocorrelation of the innovation is nearly flat and close to zero, confirming that the analog-based method effectively removes linear dependencies regardless of the signal. Moreover, the innovation kurtosis is higher that of the smallest scale increment, indicating the persistence of heavy tails in the innovation distribution. This suggests that extreme fluctuations remain largely unresolved and continue to dominate the prediction error statistics.

In the case of the GOY pseudo-velocity and the experimental velocity, we study the second order structure function of the velocity and the cumulative sum of the innovation. The second order structure function of both velocities presents the expected behavior for fully developed turbulence. In contrast, the structure function of the innovation of  $\mathsf{v}_{\text{GOY}}$ presents a slope $1$ on a log-log plot. This is the expected behavior of a random walk, and hence corresponds to an innovation signal behaving as a white noise. In the case of experimental turbulence, this linear scaling appears only for scales larger than the integral scale, indicating that the analog-based forecast does not fully remove all the linear dependencies.

Our results show that unpredictability is not uniformly distributed across scales but increases systematically toward smaller scales. It is intuitive to consider that large scale dynamics may be easily predicted at small forecast times compared to small scale dynamics. Each GOY shell has a characteristic eddy turnover time, with low frequency shells presenting larger turnover times. So, we expect the low frequency shells to be more predictable at small forecast times than the high frequency shells. However, the variation of unpredictability across scales closely mirrors the amplitude and distribution of extreme events across scales, establishing a possible link between the two: the stronger the intermittent behavior at a given scale, the higher the associated forecast error. Small-scale shells, where intermittent dynamics are strongest, are also the most difficult to forecast. The appearance of very localized peaks in both the innovation and the variance of the prediction that are concomitant to extreme values of the increments of the shell time series support this conclusion since the unpredictability of high frequency shells is not uniformly distributed in time. In this sense, unpredictability is not an external artifact, but an intrinsic feature of the cascade, tightly connected to the amplitude and distribution of extremes at each scale. This interpretation is consistent with a recent study showing that unresolved small-scale dynamics systematically increase the uncertainty of analog-based predictions~\cite{platzerEffectsUnresolvedScales2024}, highlighting the fundamental limitations of forecasting in the absence of fine-scale information, and with the idea that microscopic fluctuations may govern the selection of flow trajectories in turbulence~\cite{ruelleMicroscopicFluctuationsTurbulence1979}.

By applying a unified forecasting framework to both a synthetic model and experimental data, we provide evidence that the structure of forecast errors is defined by the physics of turbulence, and in particular its intermittent nature. This approach make it possible to assess predictability across scales, and to isolate the statistical contributions of extreme events to forecast uncertainty in turbulent systems. This uncertainty, defined in Eq.(\ref{eq:meanvariance}), depends on the pertinence of the found ensemble of analogs and of the forecast model applied on this ensemble. So, localized peaks of uncertainty imply heterogeneity of the process in terms of predictability. Concomitant extreme values of the forecast uncertainty and of the velocity increments indicate the presence of precursors of extreme events in the velocity field as well as the major unpredictability of these extreme events. These behaviors may be connected to the concept of spontaneous stochasticity, recently proposed as a mechanism for the emergence of intrinsic randomness in deterministic systems under singular limits~\cite{bandakSpontaneousStochasticityAmplifies2024,Barlet2025,Ruffenach2025}. In the context of turbulence, such randomness is hypothesized to arise from singularities in the velocity field~\cite{Drivas2021}, and in the high-Reynolds-number limit, it would imply both the non-uniqueness of solutions to the Navier–Stokes equations and the fundamental unpredictability of individual flow realizations~\cite{bandakSpontaneousStochasticityAmplifies2024}. In this context, analog-based ensembles may be a promising approach to improve our understanding of spontaneous stochasticity.

Finally, despite their similarity, see figure~\ref{fig:v_GOYsum_Modane}, we conclude that the experimental turbulence exhibits a significantly more complex structure than GOY by analyzing the average forecast performance on the experimental and GOY velocities. The pseudo-velocity signal proposed here was constructed as a real-valued combination of GOY shell variables. This reconstruction reproduces key features of turbulent velocity fields, mainly a second-order structure function consistent with the expected $ S_2(\tau) \sim \tau^{2/3} $ scaling. However, the increment also exhibits extremely high kurtosis, pointing to an over-amplification of rare, extreme fluctuations. These results indicate that while the pseudo-velocity captures the global scaling behavior of turbulence, it may distort the statistical distribution of fluctuations, particularly through nonphysical intermittency. 

In this work, we decided to study the mean and variance of the forecast pdf $q(x_t | x_{t-pdt:t-dt})$. However, from this conditional probability, other characterizations are possible, for example, predictions can be randomly sampled following $q(x_t | x_{t-pdt:t-dt})$, see~\cite{Biferale2017}. Also, in this work, we studied forecast at a fixed very small time step $dt$ corresponding to a time scale in the dissipative domain. A detailed study at different forecast times $q(x_{t} | x_{t-ps:t-s})$ could illustrate how unpredictability depends on $s$. This is specially interesting in the case of the GOY shells since each shell presents a different characteristic eddy turnover time and so we expect the shells to become completely unpredictable at time lengths of the order of the turnover times.

\appendix

\section{Construction of a pseudo velocity field from shell Models}\label{sec:appendix}

\begin{figure*}[t]
    \centering
    \includegraphics[width=16cm]{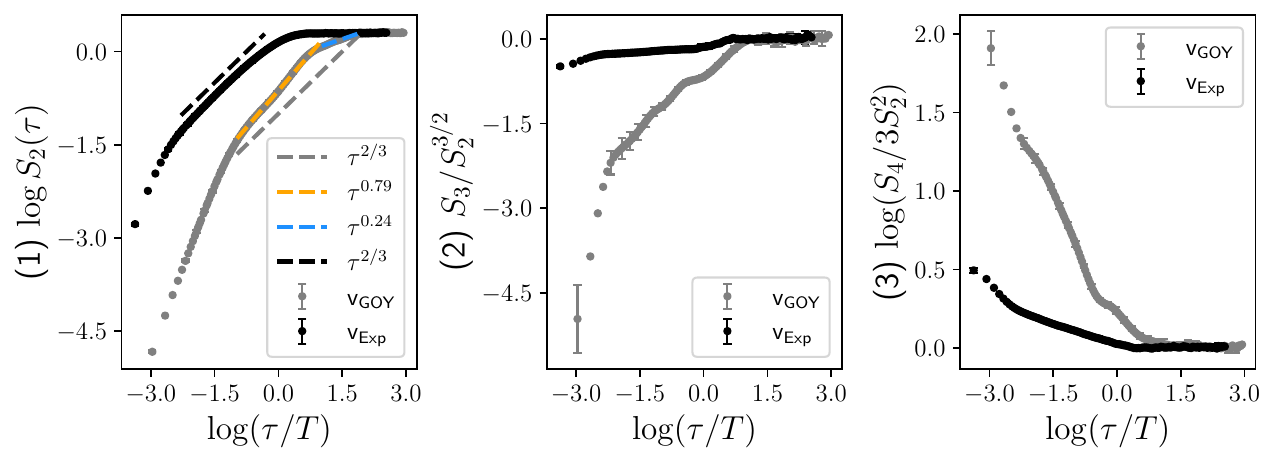}    
    \caption{Normalized structure functions of the reconstructed pseudo-velocity signal $\mathsf{v}_{\text{GOY}}$ from GOY shell variables.}
    \label{fig:StuctureFunctions}
\end{figure*}

Shell models such as GOY provide a simplified representation of turbulence through a set of coupled ordinary differential equations of complex-valued variables $\tilde{u}_n(t)$, each associated with a shell of wavenumber $k_n$. Despite originally being referred to as ``Fourier'', ``spectral'' or ``velocity'' components, these variables are not physical velocities nor do they represent direct Fourier components of a spatial velocity field.

The velocity-shell variables $\tilde{u}_n(t)$ can be understood as filtered turbulent velocities in the Fourier domain, each containing information about a limited range of scales or shell. Importantly, the GOY model is constructed to emulate the behavior of the spectral Navier--Stokes equation, but is not derived from it directly. As a result, no evident inverse transformation exists to recover a true velocity field from the \( \tilde{u}_n(t) \).

For the purpose of studying predictability and multiscale statistics, we constructed a surrogate one-dimensional pseudo-velocity signal. We define $\mathsf{v}_{\text{GOY}}(t)$ by taking the real part of the sum of all GOY shell variables, see Eq.(\ref{eq:pseudov}). 
This construction relies on the assumption that shell variables have disjoint spectral support, ensuring that each scale is not redundantly counted in the summation. Though heuristic, this approach yields a signal with multiscale characteristics and temporal dynamics resembling those of real turbulent velocity fields.

Alternative reconstructions were tested, including phase-associating the shells as Fourier components, \emph{i.e.}, assuming $\mathsf{v}_{\text{GOY}}(t) = \Re(\sum_n \tilde{u}_n(t)e^{i \omega_n t})$, with $\omega_n=k_n^{2/3}$. However, this produced strongly oscillatory signals with spectral spikes at each $\omega_n$, and did not reproduce a realistic velocity spectrum.

The reconstructed signal \( \mathsf{v}_{\text{GOY}} \), obtained by summing the real parts of the GOY variables, exhibits statistical signatures that partly align with those expected in fully developed turbulence. In particular, the second-order structure function \( S_2(\tau) \) displays a power-law scaling over a range of time scales. As shown in Figure~\ref{fig:StuctureFunctions}, panel (1), the slope of \( \log S_2(\tau) \) is approximately \( 2/3 \) over intermediate scales, consistent with Kolmogorov's prediction. However, a closer inspection reveals the presence of two distinct scaling regimes, with slopes \(\approx 0.80\) and \(\approx 0.25\), indicating a deviation from ideal inertial range behavior.

Higher-order structure functions further confirm that the pseudo-velocity is not a perfect turbulent surrogate. Figure~\ref{fig:StuctureFunctions}, panels (2) and (3), show the skewness \( S_3 / S_2^{3/2} \) and logarithm of the flatness \(\log( \frac{S_4}{3S_2^2}) \). While both quantities exhibit trends qualitatively similar to turbulence, a negative skewness and a flatness larger than 3 in both the inertial and dissipative ranges, their magnitudes and scaling behaviors differ significantly from those observed in experimental data. This discrepancy may be explained by the multiscale nature of each individual GOY shell variable. Although each $u_n$ is supposed to live in a single shell of wavenumbers, its time series contains energetic contributions across a broad range of temporal scales, as evidenced by their spectra (Figure \ref{fig:spectrum_goy}). Even at a fixed shell, the $u_n(t)$ exhibits cascade-like dynamics, with distinguishable integral, inertial, and dissipative regimes. Consequently, the pseudo-velocity signal $\mathsf{v}_{\text{GOY}}(t)$ aggregates overlapping multiscale processes. This goes against the assumption of disjoint scale contributions in spectral space, which was used to justify the way the signal was built, and may introduce artifacts in the resulting statistics.

section{Sentitivity analysis of the analog research parameters}\label{sec:appendix2}

\begin{figure*}[t]
    \centering
    \includegraphics[width=16cm]{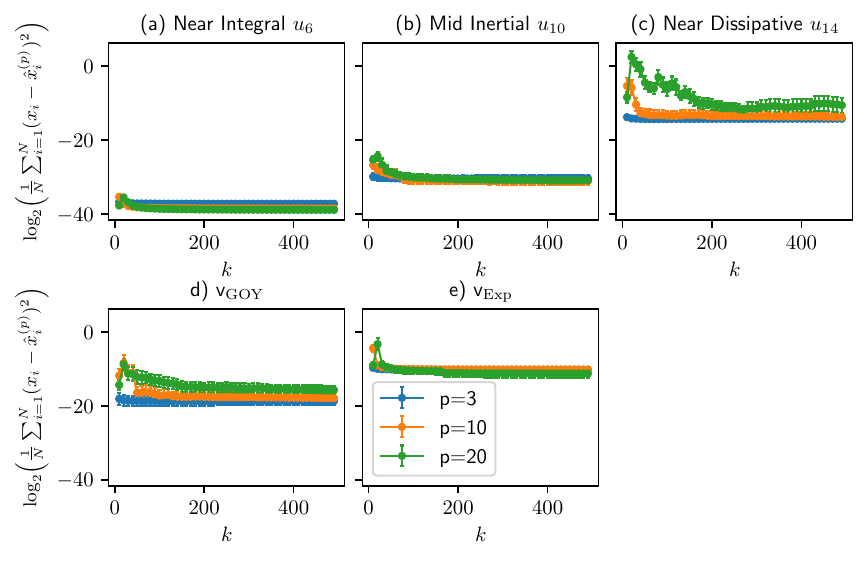}    
    \caption{Mean Squared Error $\frac{1}{N} \sum_{j=1}^{N} (x_j-\hat{x}_j^{(p)})^2$, where $x_j$ and $\hat{x}_j^{(p)}$ as a function of the number $k$ of neighbors used in the kNN algorithm. The performance is shown for $p=3,10,20$.}
    \label{fig:appB}
\end{figure*}

Figure~\ref{fig:appB} illustrates the global performance of the analog forecast in function of $k$ and $p$ for the GOY shells $u_6$, $u_{10}$ and $u_{14}$, the GOY pseudo-velocity $\mathsf{v}_{\text{GOY}}$ and the experimental turbulent velocity $\mathsf{v}_{\text{Exp}}$. The global performance is evaluated independently for each process with the mean squared error $\frac{1}{N} \sum_{j=1}^{N} (x_j-\hat{x}_j^{(p)})^2$, where $x_j$ and $\hat{x}_j^{(p)}$ are respectively the value we want to predict and the prediction, both at time $j$. The global performance is evaluated for $p=\{3,10,20\}$ and $k\in[10,500]$, using a reference database of size $N=2^{22}$ for GOY signals and $N=2^{21}$ for the experimental velocity.
 Independently of the studied process the mean squared error reaches a plateau of best performance at large $k$ indicating the minimum value of analogs needed to provide forecasts. This plateau is reached very fast for $p=3$ and slowly for larger $p$.  For the GOY shell $u_6$ and the experimental velocity $v_{\text{Exp}}$, the best performance is obtained with $p=20$, while for the other processes, which are rougher and more intermittent, it is with $p=3$.

\bibliographystyle{plainnat}

\bibliography{ArticleGOY_Innovation}
.

\end{document}